\documentclass[aps,prl,superscriptaddress,twocolumn,floatfix]{revtex4-2}

\usepackage{bm}
\usepackage{physics}
\usepackage{graphicx}
\usepackage{float}
\usepackage{multirow}
\usepackage{amsmath,amssymb,amsfonts,amsthm,mathtools}
\usepackage[colorlinks=true,linkcolor=blue,citecolor=blue,urlcolor=blue]{hyperref}
\usepackage{pdfpages}
\usepackage{amsmath}
\usepackage{amssymb}
\usepackage{tikz-cd}
\usepackage{booktabs}
\usepackage{verbatim}
\usepackage[T1]{fontenc}
\usepackage[utf8]{inputenc}
\usepackage{textcomp}
\usepackage{pgffor}

\makeatletter
\AtBeginDocument{\let\LS@rot\@undefined}
\makeatother

\begin{document}

\title{Classification of spin space groups and crystal magnetic orders}

\author{Zheng Zhang}
\affiliation{Department of Physics, School of Science, Lanzhou University of
Technology, Lanzhou 730050, China}

\author{Peiyuan Wang}
\affiliation{Department of Physics and HK Institute of Quantum Science \&
Technology, The University of Hong Kong, Pokfulam Road, Hong Kong, China}

\author{Y. X. Zhao}
\email{yuxinphy@hku.hk}
\affiliation{Department of Physics and HK Institute of Quantum Science \&
Technology, The University of Hong Kong, Pokfulam Road, Hong Kong, China}

\begin{abstract}
Spin space groups (SSGs) describe magnetic crystals in the limit of negligible spin-orbit coupling, where spatial and spin rotations decouple. Although algebraic enumerations of SSGs exist, their abstract mathematical approaches obscure the physical meaning of the resulting classifications, leaving the direct correspondence between SSGs and real-space magnetic orders unestablished. Here, we present a physically motivated framework for enumerating and classifying SSGs based on their \emph{translation holonomy}---the image of the spatial lattice translation group under the spin homomorphism. The three classes of abelian images in $SO(3)$ divide SSGs into three basic translation types. Translation holonomy constrains the spin operations associated with point-group elements, yielding a complete enumeration of SSGs organized by a physical hierarchy. Within this framework, translation holonomy classifies SSGs and magnetic orders into 17 universal spin-lattice types, each with unambiguous Bragg-peak geometries and polarization signatures in neutron diffraction. Beyond organizing known magnetic orders, this framework directly predicts four universal physical phenomena: (i) exact magnetization compensation enforced by certain spin-lattice types; (ii) an exact group-theoretic criterion for the emergence of noncommuting Brillouin zones; (iii) a new type of 3Q magnetic order with vanishing octupoles; and (iv) the realization of strictly collinear momentum-space spin textures protected by reflection-staggered coplanar spin lattices. We also construct a spin space group database organized according to this physical hierarchy.
\end{abstract}

\maketitle

\textit{\textcolor{blue}{Introduction.}} Symmetry principles underpin our understanding of quantum phases in crystalline solids~\cite{Hahn2002}. In magnetic materials, magnetic space groups (MSGs) have traditionally been used to describe symmetries that combine lattice operations with time-reversal transformations~\cite{Opechowski1965,BradleyCracknell,Litvin2013}. However, when relativistic spin-orbit coupling (SOC) is negligible, spin orientations decouple from the spatial coordinate frame, and the exact symmetry is described by a spin space group (SSG), in which spatial and spin transformations act independently~\cite{Brinkman1966,Litvin1974,Litvin1977}.

SSGs have recently emerged as an essential tool for predicting unconventional physical phenomena~\cite{Liu2022}, including altermagnetism~\cite{Smejkal2022,Smejkal2022b,Bai2024}, emergent topological excitations~\cite{Corticelli2022,Chen2025}, novel topological phases~\cite{Xiao2024,Chen2025}, and many other phenomena~\cite{Yang2024,Song2025,Feng2025,Kokkeler2025,Zhu2025,Liu2026}. Recent studies have also developed mathematical enumerations of SSGs~\cite{Xiao2024,Jiang2024,Chen2024}. In Refs.~\cite{Jiang2024,Chen2024}, SSGs are enumerated by searching for all normal subgroups of parent space groups. This approach requires a cutoff on the order of the normal subgroups, leaving the enumeration incomplete. In Ref.~\cite{Xiao2024}, a complete enumeration is achieved by exhaustively constructing the $O(3)$ representations of parent space groups by their irreducible representations. Both approaches, however, are mathematically abstract and obscure the physical information encoded in the magnetic orders associated with SSGs. Moreover, the organization of the resulting tens of thousands of SSGs is difficult to grasp intuitively. Consequently, identifying the symmetry group of a given magnetic order typically requires an exhaustive database search~\cite{Gallego2016,Chen2024,Zhang2026}.

By contrast, the 230 crystallographic space groups are organized through a layered classification built on a natural hierarchy of physical concepts~\cite{Hahn2002,Zassenhaus1948}, making the physical meaning of each space group transparent. Translation groups act on Bravais lattices; point groups and their actions on these lattices determine geometric crystal classes; and semidirect or nonsymmorphic extensions complete the classification. An analogous layered classification of SSGs, grounded in a hierarchy of physical concepts, is therefore highly desirable.

In this Letter, we develop a new enumeration method that provides such a classification framework for SSGs. Inspired by the classification of space groups, we introduce a sequence of concepts to describe how spin operations combine with translations and point-group operations. We begin with the homomorphism from the lattice translation group $\mathrm{T}$ to the spin group $O(3)$, which we call \textit{translation holonomy}. Although every space group is infinite, its quotient by the translation subgroup---the point group---is finite. Thus, once the translation holonomy is fixed, incorporating point-group elements becomes a finite problem that can be solved efficiently using numerical methods. Compatibility between the translation holonomy and the point-group action imposes a covariance condition, while an obstruction 2-cocycle determines whether consistent spin operations can be assigned to point-group elements. When this obstruction is trivial, all inequivalent SSGs can be obtained by enumerating 1-cohomology classes. At every stage of the enumeration, the mathematical objects and their relations have transparent physical meanings.

Using translation holonomy, we classify SSGs and magnetic orders in crystals into three basic translation types and 17 universal spin-lattice types (Table~\ref{tab1}). Each spin-lattice type corresponds to a specific translational pattern of magnetic order and exhibits distinct signatures in polarized neutron diffraction. We map the landscape of known magnetic orders in crystals onto this framework.

Finally, we derive four universal consequences directly from this framework: exact magnetization compensation enforced by spin-lattice type, an exact group-theoretic criterion for noncommuting Brillouin zones, a new type of 3Q magnetic order with vanishing octupoles, and symmetry protection of purely collinear spin textures in coplanar magnets.

\textit{\textcolor{blue}{Antiunitary grading and translation holonomy.}} We first formulate the framework for noncoplanar SSGs and subsequently treat coplanar and collinear SSGs as degenerate cases.

For a noncoplanar spin space group $\mathcal{G}$, a general element is written as $\{\varrho(g) | g\}$, where $g = \{R_p | \boldsymbol{t}\} \in G$ is a space-group operation and $\varrho(g) \in O(3)$ acts on the magnetic moment field $\boldsymbol{M}(\boldsymbol{r})$. Because noncoplanar SSGs possess no nontrivial spin-only operations, $\mathcal{G} \cong G$. Enumerating all inequivalent noncoplanar SSGs for a given parent space group $G$ is therefore equivalent to classifying all inequivalent homomorphisms $\varrho: G \to O(3)$.

$O(3) \cong SO(3) \times {Z}_2^\mathcal{T}$, where ${Z}_2^\mathcal{T} = \{E, \mathcal{P}\}$ represents the time-reversal group. $\mathcal{P}=\mathrm{diag}(-1,-1,-1)$ is the spin-inversion operation, realized by time reversal $\mathcal{T}$. The homomorphism $\varrho$ factors as
\begin{equation}
\varrho(g) = \chi(g) \rho(g), \quad \chi: G \to {Z}_2^\mathcal{T}, \quad \rho: G \to SO(3),
\end{equation}
where $\chi$ is a ${Z}_2$ grading specifying the unitary or antiunitary character, and $\rho$ is a homomorphism from $G$ to $SO(3)$. The inequivalent antiunitary ${Z}_2$ gradings $\chi$ are classified by $H^1(G,\mathbb{Z}_2)$. The remaining task is to enumerate all inequivalent homomorphisms $\rho: G \to SO(3)$.

%and correspond to type-I, III, IV magnetic space groups for a given OG parent group $G$~\cite{Opechowski1965,BradleyCracknell,Litvin2013}

The space group $G$ is an extension of the point group $P$ by the abelian translation subgroup $\mathrm{T}$:
\begin{equation}
1 \to \mathrm{T} \to G \to P \to 1.
\end{equation}
Restricting $\varrho$ to $\mathrm{T}$ defines the translation holonomy $\varrho_T \equiv \varrho|_\mathrm{T}$, with its $SO(3)$ part denoted by $\rho_T \equiv \rho|_\mathrm{T}$. Because $\mathrm{T}$ is abelian, its image $\rho_T(\mathrm{T})$ must be an abelian subgroup of $SO(3)$. Up to global $SO(3)$ conjugation, there are exactly three classes of abelian subgroups of $SO(3)$:

\textbf{Type 0 (Trivial):} $\rho_T(\mathrm{T}) = \{E\}$.

\textbf{Type 1 (Uniaxial):} Subgroups of $SO(2)$, generated by rotations about a fixed $z$-axis.

\textbf{Type 2 (Biaxial):} The Klein four-group $V_4 = \{E, R_x(\pi), R_y(\pi), R_z(\pi)\}$.

The translation holonomy thus classifies SSGs into three basic translation types.

\textit{\textcolor{blue}{Enumeration of noncoplanar SSGs.}} Any element of $G$ can be decomposed as $g = t s(p)$, where $s(p) = \{R_p | \boldsymbol{\tau}_p\} $ is a chosen section of the point-group operation $p \in P$ in $G$, with $\boldsymbol{\tau}_p$ a fractional translation and $\boldsymbol{t} \in \mathrm{T}$ is a pure translation. The homomorphism condition $\rho(g_1)\rho(g_2) = \rho(g_1 g_2)$ reduces to
\begin{align}
\label{hom1}
&\rho_T(\boldsymbol{t}_1)\rho_T(\boldsymbol{t}_2) = \rho_T(\boldsymbol{t}_1 + \boldsymbol{t}_2), \\
\label{hom2}
&\rho(s(p))\rho_T(\boldsymbol{t})\rho(s(p))^{-1} = \rho_T(R_{p}\boldsymbol{t}),  \\
\label{hom3}
&\rho(s(p_1))\rho(s(p_2)) = \rho_T(\boldsymbol{t}_{p_1,p_2})\rho(s(p_1p_2)), 
\end{align}
where $L$ is the Bravais lattice associated with $\mathrm{T}$, and $\boldsymbol{t}_{p_1,p_2} \in L$ is the 2-cocycle defined by $\boldsymbol{t}_{p_1,p_2}=\boldsymbol{\tau}_{p_1}+R_{p_1}\boldsymbol{\tau}_{p_2}-\boldsymbol{\tau}_{p_1p_2}$.

Because $\mathrm{T}$ is normal in $G$, $\rho_T(\mathrm{T})$ is normal in $\rho(G)$, which requires $\rho(G) \subseteq N_{SO(3)}(\rho_T(\mathrm{T}))$, where $N_{SO(3)}$ is the normalizer in $SO(3)$. We now analyze the three translation types:

\textit{Type 0.} Here $\rho_T(\mathrm{T}) = \{E\}$, so $N_{SO(3)}(\{E\}) = SO(3)$. Hence, $\rho$ is determined entirely by a homomorphism $\rho_P: P \to SO(3)$, classified by $\mathrm{Hom}(P, SO(3))$. Including the antiunitary grading $\chi$ completes the enumeration. When $\chi_T\equiv\chi|_\mathrm{T}$ is trivial, Type 0 SSGs with parent space group $G$ are in one-to-one correspondence with spin point groups associated with $P$~\cite{Litvin1977}.

\textit{Type 1.} Here $\rho_T(\mathrm{T}) \subset SO(2)$, so translation-induced spin rotations share a common $z$-axis. By Eq.~(\ref{hom1}), $\rho_T$ takes the form
\begin{equation}
\rho_T(\boldsymbol{t}) = e^{-i (\boldsymbol{k}\cdot\boldsymbol{t}) J_z},
\end{equation}
where $\boldsymbol{k} \in \mathbb{R}^3/\widehat{L} \equiv \mathrm{BZ}$ ($\boldsymbol{k} \neq \boldsymbol{0}$) represents the magnetic propagation wave vector. The normalizer of $\rho_T(\mathrm{T})$ in $SO(3)$ is $N(\rho_T(\mathrm{T})) = SO(2) \rtimes \{E, R_x(\pi)\} \cong SO(2)\rtimes C_2$. For any $s(p)\in G$, $\rho(s(p))$ decomposes as
\begin{equation}
\rho(s(p)) = e^{-i\theta(s(p))J_z} [R_x(\pi)]^{(1-\eta_p)/2},
\end{equation}
where $\eta: P \to \{\pm 1\}$ is a $\mathbb{Z}_2$ homomorphism; $\eta_p = 1$ ($-1$) indicates that $p$ preserves (reverses) the orientation of the $z$-axis. The compatibility condition in Eq.~(\ref{hom2}) requires
\begin{equation}
R_p \boldsymbol{k} = \eta_p \boldsymbol{k} \pmod{\widehat{L}}, \quad \forall p \in P.
\end{equation}
We call this equation the covariance condition for the translation holonomy. Given $\eta$, it constrains the propagation wave vector $\boldsymbol{k}$.

Furthermore, Eq.~(\ref{hom3}) requires the phase $\theta(s(p))$ assigned to each point-group section to satisfy
\begin{equation}\label{1coc}
\theta(s(p_1)) + \eta_{p_1}\theta(s(p_2)) = \boldsymbol{k}\cdot \boldsymbol{t}_{p_1,p_2} + \theta(s(p_1p_2)) \pmod{2\pi}.
\end{equation}
Thus, $\theta(s(p))$ defines a one-dimensional corepresentation with factor system $\nu_{\boldsymbol{k}}(p_1,p_2) = e^{-i\boldsymbol{k}\cdot\boldsymbol{t}_{p_1,p_2}}$. We call Eq.~(\ref{1coc}) the obstructed 1-cocycle equation for point-group elements. A valid solution requires the obstruction $\nu_{\boldsymbol{k}}\in H^{2,\eta}(P,U(1))$ to be trivial, i.e., $[\nu_{\boldsymbol{k}}] = 1$; this imposes additional selection rules on the allowed values of $\boldsymbol{k}$. Once $[\nu_{\boldsymbol{k}}] = 1$, the phases $\theta(s(p))$ are classified by $H^{1,\eta}(P, U(1))$. To enumerate all Type 1 SSGs, we determine all valid $\eta$, find the compatible wave vectors $\boldsymbol{k}$, solve for $\theta(s(p))$, and combine the results with $\chi$.

\textit{Type 2.} Here $\rho_T(\mathrm{T}) \cong V_4$. Because $V_4$ has exponent 2, translations satisfy $\rho_T(2\boldsymbol{t}) = E$, which gives
\begin{equation}
\rho_T(\boldsymbol{t}) = \mathrm{diag}\left(e^{i\boldsymbol{k}_1\cdot\boldsymbol{t}}, e^{i\boldsymbol{k}_2\cdot\boldsymbol{t}}, e^{i\boldsymbol{k}_3\cdot\boldsymbol{t}}\right),
\end{equation}
where $\boldsymbol{k}_1, \boldsymbol{k}_2, \boldsymbol{k}_3 \in \mathrm{TRIM}$ are distinct nonzero time-reversal-invariant momenta satisfying $\boldsymbol{k}_1 + \boldsymbol{k}_2 + \boldsymbol{k}_3 = \boldsymbol{0} \pmod{\widehat{L}}$. The normalizer of $V_4$ in $SO(3)$ is the chiral octahedral group $O$, with $N(V_4)/V_4 \cong S_3$, the permutation group of the three orthogonal Cartesian spin axes. Therefore, $\rho$ factorizes as $\rho(s(p)) = D(s(p)) S(p)$, where $D(s(p)) \in V_4$ and $S: P \to S_3$ is a homomorphism. By Eq.~(\ref{hom2}), the point-group action must satisfy the following covariance condition:
\begin{equation}
R_p \boldsymbol{k}_i = \boldsymbol{k}_{\sigma_p(i)} \pmod{\widehat{L}}, \quad \sigma_p \in S_3.
\end{equation}
Thus, point-group operations must permute the wave vectors.
Equation~(\ref{hom3}) then yields the obstructed 1-cocycle equation
\begin{equation}\label{2obs}
D(s(p_1))\pi_{p_1}[D(s(p_2))] = \rho_T(\boldsymbol{t}_{p_1,p_2}) D(s(p_1p_2)),
\end{equation}
where $\pi_p(D) = S(p) D S(p)^{-1}$. Equation~(\ref{2obs}) has a solution when $\rho_T(\boldsymbol{t}_{p_1,p_2})\in H^{2,\pi}(P,V_4)$ is a trivial obstruction, and inequivalent $D$ are classified by $H^{1,\pi}(P,V_4)$. Finally, combining these solutions with $\chi$ completes the enumeration of Type 2 noncoplanar SSGs. 

Equivalent SSGs must be merged during the enumeration. Details of the enumeration algorithm are given in the Supplemental Material (SM)~\cite{SM}.

\textit{\textcolor{blue}{Enumeration of coplanar and collinear SSGs.}} We now enumerate coplanar and collinear SSGs.

\textit{Coplanar SSGs.} When spins are confined to a plane, the spin-only subgroup is $\mathcal{S} = \{E, \mathcal{P}R_z(\pi)\}$. The quotient $\mathcal{G}/\mathcal{S} \cong G$ is isomorphic to a space group. The spin operation associated with $g\in G$ can be written as $\mathrm{diag}(\rho(g),\det\rho(g))$, where $\rho(g)$ is a homomorphism from $G$ to $O(2)_z$. Thus, coplanar SSGs are also classified by inequivalent homomorphisms $\rho: G\to O(2)_z\subset SO(3)$. Up to $O(2)$ conjugation, the abelian image $\rho_T(\mathrm{T})$ falls into three classes:

\textbf{Type 0:} $\rho_T(\mathrm{T}) = \{E\}$.

\textbf{Type 1:} $\rho_T(\mathrm{T}) \subset SO(2)_z$, with $\rho_T(\boldsymbol{t}) = e^{-i(\boldsymbol{k}\cdot\boldsymbol{t})J_z}$.

\textbf{Type 2:} $\rho_T(\mathrm{T}) \cong V_4 =\{E,-E,\sigma,-\sigma\}$, where $\sigma=\mathrm{diag}(1,-1)$, or $\rho_T(\mathrm{T}) \cong D_1=\{1,\sigma\}$. Their embeddings in $SO(3)$ are $V_4=\{E, R_x(\pi), R_y(\pi), R_z(\pi)\}$ and $D_1 = \{E, R_x(\pi)\}$.

Type 0 coplanar SSGs correspond to coplanar spin point groups. Type 1 coplanar SSGs are in one-to-one correspondence with Type 1 noncoplanar SSGs carrying a trivial antiunitary grading. For Type 2 coplanar SSGs, we write $\rho_T(\boldsymbol{t}) = \mathrm{diag}(e^{i\boldsymbol{k}_1\cdot \boldsymbol{t}}, e^{i\boldsymbol{k}_2\cdot \boldsymbol{t}})$, where $\boldsymbol{k}_1, \boldsymbol{k}_2$ are distinct TRIMs; both are nonzero when $\rho_T(\mathrm{T}) \cong V_4$, while one is zero when $\rho_T(\mathrm{T}) \cong D_1$. Type 2 coplanar SSGs can be enumerated using a method analogous to that for the noncoplanar case; details are given in the SM.

\textit{Collinear SSGs.} When moments align along a single axis, the spin-only group is $\mathcal{S} = SO(2) \rtimes \{E, \mathcal{P}R_x(\pi)\}$, and $\mathcal{G}/\mathcal{S} \cong G$. Collinear SSGs are determined by homomorphisms $\varrho: G \to O(1)=\{E,\mathcal{P}\}$, which are precisely the antiunitary ${Z}_2$ gradings $\chi$ of $G$ (corresponding to Type I, III, and IV MSGs). Within our holonomy framework, all collinear SSGs belong to Type 0.

\textit{\textcolor{blue}{Spin lattice types.}}
Space groups and crystal orders can be classified into 14 Bravais types according to their lattice symmetries. For magnetic orders in crystals, each lattice site is associated with a set of magnetic moments or spins; we call the resulting structure a spin lattice. Similarly, the translation holonomy image $\varrho_T(\mathrm{T})$ classifies SSGs and magnetic orders into 17 universal spin-lattice types (Table~\ref{tab1}). In the following, we characterize the antiunitary grading $\chi_T$ by a time-reversal-invariant wave vector $\boldsymbol{k}_\chi$: $\chi_T(\boldsymbol{t})=e^{i\boldsymbol{k}_\chi\cdot\boldsymbol{t}}$. %Our framework clarifies the previous classification of magnetic orders in the language of symmetry, and also reveals new types not previously identified. %Here we note our classification should not be confused with conventional classification of magnetic orders via propagation vectors, which is not grounded in symmetry group, while our classification is grounded in the complete classification of SSGs. 

\begin{table*}[t]
\caption{The 17 spin-lattice types of spin space groups, classified by translation holonomy $\varrho_T(\mathrm{T}) \subset O(n)$ and propagation wave vectors. Here, $\boldsymbol{k}_\chi$ governs the antiunitary translation grading $\chi_T$. In this table, TRIM denotes a nonzero time-reversal-invariant momentum. For Type 2 noncoplanar systems, $\boldsymbol{k}_i$ ($i=1,2,3$) are three nonzero TRIMs satisfying $\sum_i \boldsymbol{k}_i = \boldsymbol{0}$.}
\label{tab1}
\centering
\footnotesize
\setlength{\tabcolsep}{4pt}
\begin{tabular}{@{}llll@{}}
\toprule
Type & Wave Vectors & Spin Lattice Type & Fourier Wave Vectors and Polarizations \\
\midrule
\multicolumn{4}{l}{\textit{Collinear}} \\
Type 0 & $\boldsymbol{k}_\chi = \boldsymbol{0}$ & L.0.1 Translation-invariant collinear & Linear along unique spin axis at $\Gamma$ \\
       & $\boldsymbol{k}_\chi \neq \boldsymbol{0}$ & L.0.2 Staggered collinear & Linear along unique spin axis at $\boldsymbol{k}_\chi$ \\
\midrule
\multicolumn{4}{l}{\textit{Coplanar}} \\
Type 0 & $\boldsymbol{k} = \boldsymbol{0}$ & P.0 Translation-invariant coplanar & In-plane at $\Gamma$ \\
Type 1 & $\boldsymbol{k} \in \mathrm{TRIM}$ & P.1.1 Staggered coplanar & In-plane at $\boldsymbol{k}$ \\
       & $\boldsymbol{k}\ \mathrm{general}$ & P.1.2 Planar spiral & Circular with opposite chirality at $\pm\boldsymbol{k}$ \\
Type 2 & $\boldsymbol{k}_1=0, \boldsymbol{k}_2 \in \mathrm{TRIM}$ & P.2.1 Reflection-staggered coplanar & Two orthogonal linear polarizations at $\Gamma$ and $\boldsymbol{k}_2$ \\
       & $\boldsymbol{k}_1, \boldsymbol{k}_2 \in \mathrm{TRIM}$ & P.2.2 Biaxial-staggered coplanar & Two orthogonal linear polarizations at $\boldsymbol{k}_1$ and $\boldsymbol{k}_2$ \\
\midrule
\multicolumn{4}{l}{\textit{Noncoplanar}} \\
Type 0 & $\boldsymbol{k}_\chi = \boldsymbol{0}$ & N.0.1 Translation-invariant noncoplanar & Unrestricted 3D polarization at $\Gamma$ \\
       & $\boldsymbol{k}_\chi \neq \boldsymbol{0}$ & N.0.2 Staggered noncoplanar & Unrestricted 3D polarization at $\boldsymbol{k}_\chi$ \\
Type 1 & $\boldsymbol{k}_\chi = \boldsymbol{0}, \boldsymbol{k} \in \mathrm{TRIM}$ & N.1.1 Bipartite conical & Longitudinal at $\Gamma$; transverse at $\boldsymbol{k}$ \\
       & $\boldsymbol{k}_\chi = \boldsymbol{0}, \boldsymbol{k}\ \mathrm{general}$ & N.1.2 Conical spiral & Linear along axis at $\Gamma$; circular at $\pm \boldsymbol{k}$ \\
       & $\boldsymbol{k}_\chi = \boldsymbol{k} \in \mathrm{TRIM}$ & N.1.3 Reflection-staggered noncoplanar & Longitudinal at $\boldsymbol{k}$; transverse at $\Gamma$ \\
       & $\boldsymbol{k} \in \mathrm{TRIM}, \boldsymbol{k}_\chi\not \in\{\boldsymbol{0}, \boldsymbol{k}\}$ & N.1.4 Alternating bipartite conical & Longitudinal at $\boldsymbol{k}_\chi$; transverse at $\boldsymbol{k}_\chi + \boldsymbol{k}$ \\
       & $\boldsymbol{k}_\chi \neq \boldsymbol{0}, \boldsymbol{k}\ \mathrm{general}$ & N.1.5 Antiferromagnetic conical spiral & Linear along axis at $\boldsymbol{k}_\chi$; circular at $\boldsymbol{k}_\chi\pm \boldsymbol{k}$ \\
Type 2 & $\boldsymbol{k}_\chi = \boldsymbol{0}, \boldsymbol{k}_i \in \mathrm{TRIM}$ & N.2.1 Orthogonal 3Q noncoplanar & Three orthogonal linear polarizations at $\boldsymbol{k}_i$ \\
       & $\boldsymbol{k}_i \in \mathrm{TRIM}, \boldsymbol{k}_\chi \in \{\boldsymbol{k}_i\}$ & N.2.2 Canted 2Q noncoplanar & Three orthogonal linear polarizations at $\boldsymbol{k}_\chi + \boldsymbol{k}_i$ (one is $\Gamma$) \\
       & $\boldsymbol{k}_i \in \mathrm{TRIM}, \boldsymbol{k}_\chi \notin \{\boldsymbol{0},\boldsymbol{k}_i\}$ & N.2.3 Staggered orthogonal 3Q noncoplanar & Three orthogonal linear polarizations at $\boldsymbol{k}_\chi + \boldsymbol{k}_i$ \\
\bottomrule
\end{tabular}
\end{table*}

\textit{Collinear spin lattices.} For trivial $\chi_T$ ($\boldsymbol{k}_\chi = \boldsymbol{0}$), spins form a translation-invariant collinear lattice, encompassing ferromagnets and ferrimagnets~\cite{Coey2010}, as well as collinear altermagnets~\cite{Smejkal2022} with crystallographic magnetic unit cells. For nontrivial $\chi_T$ ($\boldsymbol{k}_\chi \neq \boldsymbol{0}$), antiunitary translations reverse the spins, defining staggered collinear spin lattices, as realized in type-II antiferromagnets such as NiO~\cite{Roth1958}.

\textit{Coplanar spin lattices.} Coplanar configurations do not admit an independent antiunitary translation grading $\chi_T$. Type 0 gives translation-invariant coplanar lattices, as found in triangular and kagome $120^\circ$ antiferromagnets with magnetic unit cells identical to their crystallographic cells (e.g., $\mathrm{Mn_3Sn}$, $\mathrm{Mn_3Ge}$, and $\mathrm{LuMnO_3}$)~\cite{Nakatsuji2015,Cederholm2026,Soh2020,Sim2016}. 

Type 1 coplanar SSGs have $\rho_T(\mathrm{T})\subset SO(2)$ with normalizer $SO(2)\rtimes C_2$. However, $C_2$ can act on $\rho_T(\mathrm{T})$ in two inequivalent ways. When the wave vector $\boldsymbol{k}$ is a TRIM, $C_2$ acts trivially on $\rho_T(\mathrm{T})$. When $\boldsymbol{k}$ is a general point, $C_2$ acts nontrivially. We therefore divide Type 1 spin lattices into two translation types. We call the former, with TRIM $\boldsymbol{k}$, the staggered coplanar spin lattice because certain translations reverse the spins. We call the latter, with general $\boldsymbol{k}$, the planar spiral spin lattice. For a planar spiral spin lattice, $\boldsymbol{k}$ can be commensurate or incommensurate, corresponding to commensurate or incommensurate magnetic order, respectively. Representative states include proper planar helices and cycloids, such as the cycloid in $\mathrm{BiFeO_3}$~\cite{Sosnowska1982,Ramazanoglu2011} or helices in elemental rare-earth metals (Dy, Ho)~\cite{Jensen1991,Gibbs1985}.

Type 2 coplanar SSGs are divided into two translation types according to $\rho_T(\mathrm{T})$. When $\rho_T(\mathrm{T})=D_1$, one of $\boldsymbol{k}_1,\boldsymbol{k}_2$ is zero, so the nontrivial $\rho_T(\boldsymbol{t})$ acts as a reflection. We call this type the reflection-staggered coplanar spin lattice. Examples include EuTe in the canted type-II antiferromagnetic phase~\cite{Oliveira1972}. When $\rho_T(\mathrm{T})=V_4$, $\boldsymbol{k}_1,\boldsymbol{k}_2$ are distinct nonzero TRIMs, so the two spin components are staggered in different directions. We call this type the biaxial-staggered coplanar spin lattice. Examples include $\mathrm{La}_2\mathrm{O}_2\mathrm{Fe}_2\mathrm{O}\mathrm{Se}_2$ in the noncollinear two-$\boldsymbol{k}$ phase~\cite{McCabe2014} and Ni- and Co-substituted CaKF$\mathrm{e}_4$A$\mathrm{s}_4$ with hedgehog spin-vortex-crystal magnetic order~\cite{Meier2018}. 

\textit{Noncoplanar spin lattices.} The grading $\chi_T$ divides Type 0 into translation-invariant ($\boldsymbol{k}_\chi = \boldsymbol{0}$) and staggered ($\boldsymbol{k}_\chi \neq \boldsymbol{0}$) noncoplanar spin lattices. The former is realized in all-in-all-out pyrochlores~\cite{Sagayama2013,Guo2016}. One example of the latter is KF$\mathrm{e}_3\mathrm{(OH)}_6\mathrm{(SO_4)_2}$ with antiferromagnetically stacked umbrella order~\cite{Inami2000,Bartlett2005}.

For Type 1 noncoplanar spin lattices with $\boldsymbol{k}_\chi=\boldsymbol{0}$, two spin-lattice types arise, as in the coplanar case. When $\boldsymbol{k}$ is a TRIM, some translations are associated with a $\pi$-rotation about an axis; we call this the bipartite conical spin lattice. When $\boldsymbol{k}$ is a general point, spins at different sites trace a spiral on a conical surface, defining the conical spiral lattice. Representative examples of this type are $\mathrm{MnSi}$ and FeGe in their field-induced conical phase~\cite{Ishikawa1976,Lebech1989}. Unlike the coplanar case, Type 1 noncoplanar spin lattices can have $\boldsymbol{k}_\chi\not=\boldsymbol{0}$, which yields three additional spin-lattice types. When $\boldsymbol{k}_\chi=\boldsymbol{k}$, the modulations of the antiunitary grading and the $\pi$-rotation coincide, producing a modulation of the mirror reflection along the axis. We call this type the reflection-staggered noncoplanar spin lattice. When $\boldsymbol{k}_\chi$ and $\boldsymbol{k}$ are two distinct TRIMs, we call the lattice the alternating bipartite conical spin lattice. Finally, when $\boldsymbol{k}_\chi\not=\boldsymbol{0}$ and $\boldsymbol{k}$ is general, spins spiral in the transverse plane while their longitudinal components alternate, producing antiferromagnetic order. We call it the antiferromagnetic conical spiral lattice. For conical and antiferromagnetic conical spiral spin lattices, $\boldsymbol{k}$ can be commensurate or incommensurate.

For Type 2, trivial $\chi_T$ yields orthogonal 3Q noncoplanar lattices, in which three orthogonal spin components are modulated independently by $\boldsymbol{k}_1, \boldsymbol{k}_2, \boldsymbol{k}_3$. Representative systems include triple-$q$ ($3Q$) tetrahedral noncoplanar states in FCC and Kondo lattices (e.g., $\mathrm{UO}_2$ with transverse 3-$\boldsymbol{k}$ order~\cite{Laskowski2004} and USb in the 3-$\boldsymbol{k}$ antiferromagnetic phase~\cite{RossatMignod1980}). When $\boldsymbol{k}_\chi \neq \boldsymbol{0}$, two further types arise: canted 2Q noncoplanar lattices ($\boldsymbol{k}_\chi \in \{\boldsymbol{k}_i\}$, leaving one axis unmodulated) and staggered orthogonal 3Q noncoplanar lattices ($\boldsymbol{k}_\chi \notin \{\boldsymbol{k}_i\}$).

\textit{\textcolor{blue}{Neutron diffraction signatures.}}
The real-space magnetization density satisfies the translation covariance condition:
\begin{equation}\label{tranco}
\boldsymbol{M}(\boldsymbol{r} + \boldsymbol{t}) = \varrho_T(\boldsymbol{t})\boldsymbol{M}(\boldsymbol{r}).
\end{equation}
The Fourier expansion $\boldsymbol{M}(\boldsymbol{r}) = \sum_{i} \boldsymbol{M}_{\boldsymbol{Q}_i}(\boldsymbol{r}) e^{i\boldsymbol{Q}_i\cdot\boldsymbol{r}}$ shows that the propagation wave vectors $\boldsymbol{Q}_i$ and the corresponding spin polarization vectors are fixed by the eigensystem of $\varrho_T$. As detailed in Table~\ref{tab1}, each of the 17 spin-lattice types has a unique combination of Bragg-peak positions $\boldsymbol{Q}_i + \boldsymbol{G}$ and polarization profiles. Thus, polarized neutron diffraction measurements provide a direct experimental determination of the underlying spin-lattice type. %Eq.~(\ref{tranco}) also bridges symmetry descriptions of SSGs and Bertaut theory~\cite{Bertaut1968} used in magnetic structure analysis, which we discuss in the SM.

\textit{\textcolor{blue}{Universal physical consequences.}} Our classification framework yields many universal physical consequences across diverse magnetic orders. Here we highlight four examples; complete derivations are provided in the SM~\cite{SM}.

\textit{Exact compensation of magnetization.} For the spatially averaged magnetization $\overline{\boldsymbol{M}}$, translation covariance implies $\overline{\boldsymbol{M}}=\varrho_T(\boldsymbol{t})\overline{\boldsymbol{M}}$ for any $\boldsymbol{t}\in L$. Thus, the translation holonomy $\varrho$ constrains the possible uniform magnetization $\overline{\boldsymbol{M}}$. In our classification framework, nine spin-lattice types require $\overline{\boldsymbol{M}}=0$, enforcing exact compensation of the uniform magnetization. They are L.0.2, P.1.1, P.1.2, P.2.2, N.0.2, N.1.4, N.1.5, N.2.1, and N.2.3. A full analysis of the constraints on magnetization for all 17 spin-lattice types is presented in the SM.%In addition, P.2.1 constrains uniform magnetization along the intersection of the mirror plane and the spin rotation plane. N.1.1, N.1.2 constrain uniform magnetization along the spin rotation axis. N.1.3 constrains uniform magnetization in the transverse plane. N.2.2 constrains uniform magnetization along the unmodulated spin direction. 

\textit{Exact criterion for noncommuting Brillouin zones.} Lifting $\rho_T(\mathrm{T})$ to the spinor double cover $SU(2)$ can produce a nonabelian translation algebra for spin-$1/2$ electrons~\cite{Xiao2024}. Because $U(1)$ lifts remain abelian, this phenomenon occurs exclusively in Type 2 SSGs, where the lift of $V_4$ forms the quaternion group $Q_8$. In general, crystal momentum and the Brillouin zone (BZ) are defined with respect to the unitary translation subgroup $\mathrm{T}_u \equiv \ker(\chi|_\mathrm{T})$. A noncommuting BZ requires $\rho_T(\mathrm{T}_u) \cong V_4$. This yields an exact criterion:

\textit{An SSG hosts a noncommuting BZ if and only if its spin lattice is of the biaxial-staggered coplanar type (P.2.2), the orthogonal 3Q noncoplanar type (N.2.1), or the staggered orthogonal 3Q noncoplanar type (N.2.3).}

%\textit{Nonsymmorphic symmetries in momentum space.} Projective representations arising from the $SU(2)$ double cover can induce fractional momentum-space translations $\boldsymbol{\kappa}_p$:
%\begin{equation}
%\tilde{\rho}(s(p))^{-1} \tilde{\rho}_T(\boldsymbol{t}) \tilde{\rho}(s(p)) = e^{i\boldsymbol{\kappa}_p \cdot \boldsymbol{t}} \tilde{\rho}_T(R_p^{-1}\boldsymbol{t}), \quad \boldsymbol{t} \in L_u,
%\end{equation}
%where $\widetilde{\rho}$ is the SU(2) lift of $\rho$. This shifts crystal momentum according to $p: \boldsymbol{k} \to (-1)^{s_p}(R_p \boldsymbol{k} + \boldsymbol{\kappa}_p)$, with $s_p = 0$ ($1$) for unitary (antiunitary) $p$. We find that only Type 1 SSGs with non-trivial $\eta$ can host such essential nonsymmorphic momentum symmetries, where the fractional translations are given by 
%\begin{equation}
%\boldsymbol{\kappa}_p=\frac{1}{2}(R_p\boldsymbol{k}-\eta_p \boldsymbol{k}) \ \ \mod \widehat{L}_u.
%\end{equation}

\textit{3Q magnetic orders with zero octupoles.} Our classification framework predicts a new type of 3Q magnetic order, the staggered orthogonal 3Q noncoplanar spin lattice (N.2.3). Although orders of this type are superficially similar to ordinary orthogonal 3Q noncoplanar orders (N.2.1), their nonlinear magnetic responses differ. The orthogonal 3Q noncoplanar lattice allows a nonzero magnetic octupole component $\mathcal{O}_{xyz}=\overline{M_xM_yM_z}$, while the staggered orthogonal 3Q noncoplanar lattice forces the uniform octupole to vanish. Thus, these two types of 3Q magnetic order can be distinguished by second-order magnetization.

\textit{Coplanar magnets with collinear spin textures.} While coplanar magnets have noncollinear moments in real space, recent studies of coplanar odd/even-wave magnetism~\cite{Hellenes2023,Luo2025,Song2026} have identified states exhibiting strictly collinear momentum-space spin textures $\boldsymbol{S}(\boldsymbol{k})$. We show that this is a universal symmetry property of SSGs with the reflection-staggered coplanar spin lattice (P.2.1). For a nondegenerate band in this type, a translation $\boldsymbol{t}$ with spin operation $R_x(\pi)$ acts on $\boldsymbol{S}(\boldsymbol{k})$, requiring $\boldsymbol{S}(\boldsymbol{k}) = R_x(\pi) \boldsymbol{S}(\boldsymbol{k})$. This forces $S_y, S_z$ to vanish. The spin-only group forces $S_x(\boldsymbol{k})=S_x(-\boldsymbol{k})$. Thus, the spin texture is strictly collinear and has even parity. The constraints imposed on spin textures by all 17 spin-lattice types are analyzed in the SM. %Note that if there is another $\pi$-rotation $R_y(\pi)$ associated with a translation element, $\boldsymbol{S}(\boldsymbol{k})$ will be forced to be zero. 

\textit{\textcolor{blue}{Discussion and conclusion.}} We have presented a complete, layered classification scheme for SSGs with a clear physical interpretation in terms of magnetic orders. The scheme uses translation holonomy to organize SSGs into 17 spin-lattice types. Each type displays a distinct diffraction signature in momentum and polarization spaces, bridging abstract group theory directly to neutron diffraction experiments. This connection could provide a practical SSG framework for neutron diffraction refinement.

%In summary, we have established a physically transparent classification framework for spin space groups and crystal magnetic orders grounded in translation holonomy. By categorizing the image of lattice translations in spin space, SSGs are organized into three basic translation types and 17 universal spin-lattice types. Each type displays a distinct diffraction signature in momentum and polarization spaces, bridging abstract group theory directly to neutron diffraction experiments. Compared with previous enumeration and classification frameworks of SSGs~\cite{Xiao2024,Jiang2024,Chen2024,Liu2026}, our framework is more physical, understandable and efficient in applications. It can also potentially lead to a workable SSG framework for neutron diffraction refinement. %which deserves to be developed in the future.  %We compare our framework with previous works and discuss their relationships in the SM.

Beyond translation holonomy, our framework supports finer classification at several levels, as described in the End Matter. We have constructed an SSG database based on this hierarchical classification framework~\cite{SSGdatabase}. These finer levels can also provide physical information about magnetic orders; we leave their investigation to future work.

%\textbf{Acknowledgments}

\begin{acknowledgments}
\textit{\textcolor{blue}{Acknowledgments}} Claude (Opus 5.5) assisted in deriving the enumeration algorithm. GPT (6 Astra, 5.6 Sol, 6.1 Sol) assisted in deriving the physical consequences, providing material examples, and verifying the derivations throughout this work. The code and database were generated by GPT (6 Astra, 6.1 Sol) under the authors’ instructions. Gemini (3.8 Flash) suggested some of the terminology used in this work. Gemini (3.8 Flash) and DeepSeek (4.1 Flash) assisted in polishing the writing.
\end{acknowledgments}

\vspace{0.5cm}

{\centering\textbf{End Matter}\par
}

\textit{\textcolor{blue}{Appendix A: Hierarchical classification of spin space groups.}} Our enumeration algorithm provides a natural hierarchical classification framework for SSGs. For a space group $G$ with translation group $\mathrm{T}$ and point group $P$, the homomorphism $\varrho: G\to O(n)$ induces the following morphism between the two group extensions:
\[
\begin{tikzcd}
1 \arrow[r] & \mathrm{T} \arrow[r] \arrow[d,"\varrho_{T}"]
            & G \arrow[r] \arrow[d,"\varrho"]
            & P \arrow[r] \arrow[d,"\varrho_P"]
            & 1 \\
1 \arrow[r] & \varrho_{{T}}(\mathrm{T}) \arrow[r]
            & \varrho(G) \arrow[r]
            & \varrho_P(P) \arrow[r]
            & 1
\end{tikzcd}
\]
where $\varrho(G)\subset N_{O(n)}(\varrho_T(\mathrm{T}))$, $ \varrho_P(P)=\varrho(G)/\varrho_T(\mathrm{T})$.

We can now define the following hierarchy of SSGs. First, the lattice group $\mathrm{T}$ defines a lattice $L$ in $\mathbb{R}^3$, which can be classified into 14 Bravais lattice types. The translation symmetry of a spin space group is determined by $\mathrm{T}$ and $\varrho_T$. Together, $\mathrm{T}$ and $\varrho_T$ specify a lattice with spins associated with each site, which we call a spin lattice. The spatial symmetries of spin lattices can still be classified into 14 Bravais lattice types. To capture the spin symmetries of a spin lattice more fully, we introduce two new concepts: the translation holonomy type and the spin Bravais lattice type. The former is determined by $\varrho_T(\mathrm{T})$, while the latter is determined by $(\mathrm{T},\varrho_T)$.

\textbf{Translation holonomy type}: \textit{Two spin lattices belong to the same translation holonomy type if and only if they have equivalent $\varrho_T(\mathrm{T})$.}

\textbf{Spin Bravais lattice type}: \textit{Two spin lattices belong to the same spin Bravais lattice type if and only if they belong to the same Bravais lattice type and have equivalent $\varrho_T$.}

Because collinear, coplanar, and noncoplanar SSGs have different spin-only groups, their spin lattices are always inequivalent and are treated separately. 

This definition yields infinitely many translation holonomy types. For convenience, we group certain infinite sets of these types into families (see the SM for details), thereby defining the concept of a \textbf{spin-lattice type}. This Letter defines 17 such types. 

We next include point-group symmetry. At the level of spatial symmetry, $\mathrm{T}$ and $P$ together determine the arithmetic crystal class. For spin space groups, we define two additional concepts:

\textbf{Spin crystal class:} Determined by $(\varrho_T(\mathrm{T}),\varrho_P(P))$. Two SSGs belong to the same spin crystal class if and only if they have equivalent $\varrho_T(\mathrm{T})$ and equivalent $\varrho_P(P)$.

\textbf{Spin arithmetic crystal class:} Determined by $(\mathrm{T}, P, \varrho_T, \varrho_P)$. Two SSGs belong to the same spin arithmetic crystal class if and only if they belong to the same arithmetic crystal class and have equivalent $\varrho_T$ and $\varrho_P$.

In defining a spin crystal class, inequivalent embeddings of $\varrho_P(P)$ into $N_{O(n)}(\varrho_T(\mathrm{T}))/\varrho_T(\mathrm{T})$ should be distinguished. 

The concepts of spin lattice type (or translation holonomy type) and spin crystal class classify SSGs solely within the spin-operation space $O(n)$, without reference to the underlying crystal symmetry. Accounting for how spin operations are associated with spatial operations refines spin-lattice types into spin Bravais lattice types and spin crystal classes into spin arithmetic crystal classes.

\textit{\textcolor{blue}{Appendix B: Spin space group database.}} We enumerate SSGs using the algorithm presented in this Letter. Adopting the same equivalence conditions as in Ref.~\cite{Xiao2024}, we obtain 1,421 collinear, 9,542 coplanar, and 56,512 noncoplanar SSG entries. Of the 67,475 SSG entries, 58,955 are discrete groups and 8,520 are continuous families, in agreement with Ref.~\cite{Xiao2024}. We tabulate all SSGs in a database~\cite{SSGdatabase}, organized into five levels for retrieval:

1. Parent group $G$.

2. Spin configuration (collinear, coplanar, or noncoplanar).

3. Spin lattice type.

4. Spin Bravais lattice type. 

5. Spin crystal class. 

The database also supports conditional searches for SSGs. 

The database uses an SSG naming scheme similar to that of Ref.~\cite{Chen2024}, with a one-to-one correspondence to the labels in Ref.~\cite{Xiao2024}. Classification information and explicit matrices are tabulated for each SSG.

\newpage

\clearpage
\onecolumngrid
\foreach \x in {1,...,44}
{
    \clearpage
    \includepdf[pages={\x}]{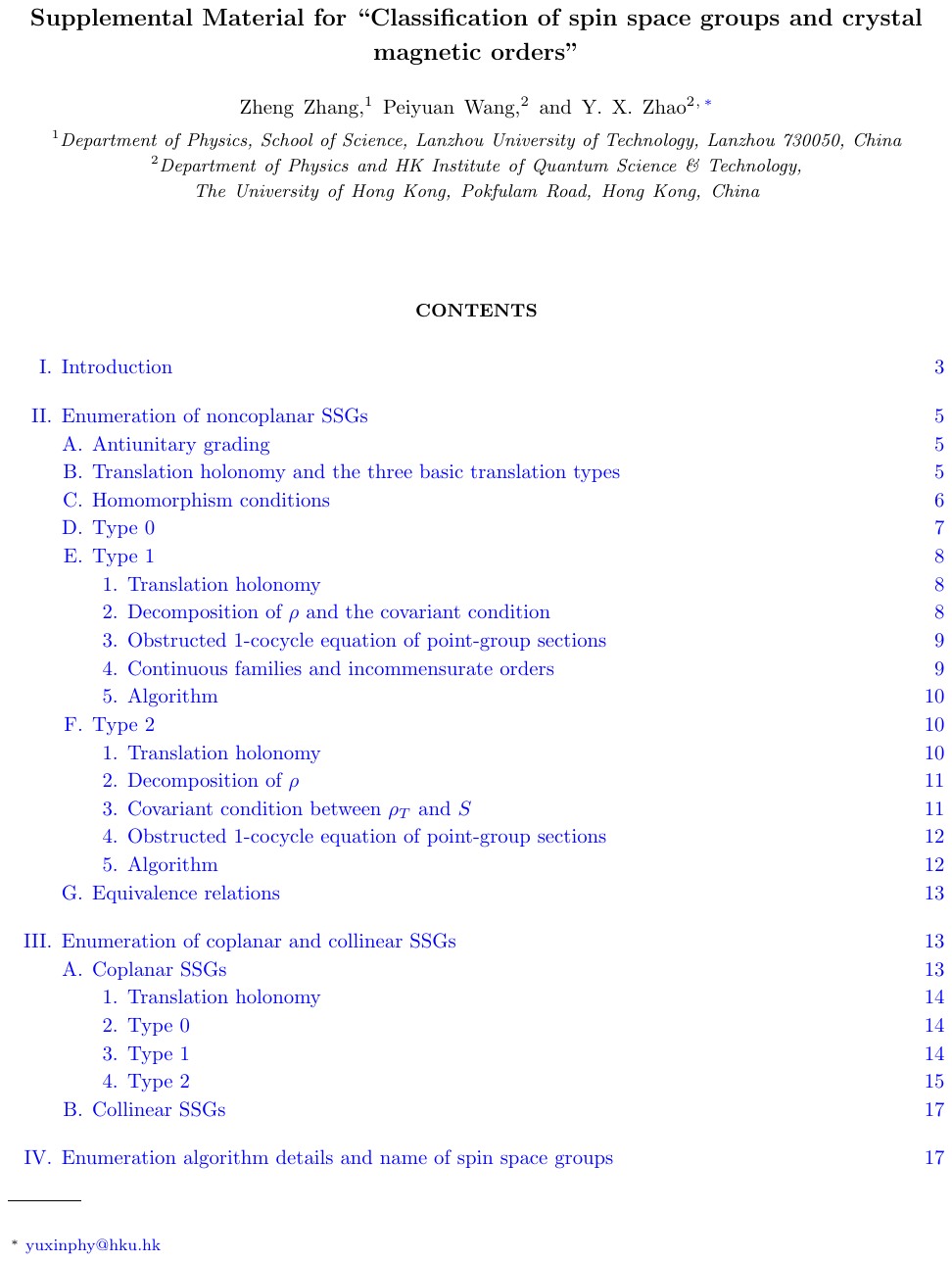}
}

\end{document}